\input harvmac\skip0=\baselineskip

\newcount\figno
\figno=0
\def\fig#1#2#3{
\par\begingroup\parindent=0pt\leftskip=1cm\rightskip=1cm\parindent=0pt
\baselineskip=11pt \global\advance\figno by 1 \midinsert
\epsfxsize=#3 \centerline{\epsfbox{#2}} \vskip 12pt {\bf Fig.\
\the\figno: } #1\par
\endinsert\endgroup\par
}
\def\figlabel#1{\xdef#1{\the\figno}}
\def\encadremath#1{\vbox{\hrule\hbox{\vrule\kern8pt\vbox{\kern8pt
\hbox{$\displaystyle #1$}\kern8pt} \kern8pt\vrule}\hrule}}

% approximately less than

% approximately greater than
%\OkounkovSP
%\MooreFG
\lref\MooreFG{
  G.~W.~Moore,
  ``Les Houches lectures on strings and arithmetic,''
  arXiv:hep-th/0401049.
  %%CITATION = HEP-TH 0401049;%%
}

%\MoorePN
\lref\MoorePN{
  G.~W.~Moore,
  ``Arithmetic and attractors,''
  arXiv:hep-th/9807087.
  %%CITATION = HEP-TH 9807087;%%
}

%\MooreZU
\lref\MooreZU{
  G.~W.~Moore,
  ``Attractors and arithmetic,''
  arXiv:hep-th/9807056.
  %%CITATION = HEP-TH 9807056;%%
}

\lref\aes{ A.~Strominger, ``Macroscopic Entropy of $N=2$ Extremal
Black Holes,'' Phys.\ Lett.\ B {\bf 383}, 39 (1996)
[arXiv:hep-th/9602111].
%%CITATION = HEP-TH 9602111;%%
}

\lref\denefb{
  F.~Denef,
  ``Supergravity flows and D-brane stability,''
  JHEP {\bf 0008}, 050 (2000)
  [arXiv:hep-th/0005049].
}

\lref\denefc{F.~Denef, B.~R.~Greene and M.~Raugas,
  ``Split attractor flows and the spectrum of BPS D-branes on the quintic,''
  JHEP {\bf 0105}, 012 (2001)
  [arXiv:hep-th/0101135].}

\lref\ascv{ A.~Strominger and C.~Vafa, ``Microscopic Origin of the
Bekenstein-Hawking Entropy", Phys.\ Lett.\ B {\bf 379}, 99 (1996)
[arXiv:hep-th/9601029]. }

\lref\BatesVX{
  B.~Bates and F.~Denef,
  ``Exact solutions for supersymmetric stationary black hole composites,''
  arXiv:hep-th/0304094.
  %%CITATION = HEP-TH 0304094;%%
}

\lref\allfive{
  J.~P.~Gauntlett, J.~B.~Gutowski, C.~M.~Hull, S.~Pakis and H.~S.~Reall,
  ``All supersymmetric solutions of minimal supergravity in five dimensions,''
  Class.\ Quant.\ Grav.\  {\bf 20}, 4587 (2003)
  [arXiv:hep-th/0209114].
  %%CITATION = HEP-TH 0209114;%%
}

\lref\bmpv{ J.C. Breckenridge, R.C. Myers, A.W. Peet and C. Vafa,
``D-branes and Spinning Black Holes", hep-th/9602065. }

\lref\CyrierHJ{
  M.~Cyrier, M.~Guica, D.~Mateos and A.~Strominger,
  ``Microscopic entropy of the black ring,''
  arXiv:hep-th/0411187.
  %%CITATION = HEP-TH 0411187;%%
}

\lref\bpsb{ M. Marino, R. Minasian, G. Moore and A. Strominger,
``Nonlinear Instantons from Supersymmetric $p$-Branes",
hep-th/9911206. }

\lref\GutowskiYV{
  J.~B.~Gutowski and H.~S.~Reall,
  ``General supersymmetric AdS(5) black holes,''
  JHEP {\bf 0404}, 048 (2004)
  [arXiv:hep-th/0401129].
  %%CITATION = HEP-TH 0401129;%%
}

\lref\DabhBY{
  A.~Dabholkar, F.~Denef, G.~W.~Moore and B.~Pioline,
  ``Exact and asymptotic degeneracies of small black holes,''
  arXiv:hep-th/0502157.
}

\lref\fks{ S.~Ferrara, R.~Kallosh and A.~Strominger, ``N=2
extremal black holes,'' Phys.\ Rev.\ D {\bf 52}, 5412 (1995)
[arXiv:hep-th/9508072].
%%CITATION = HEP-TH 9508072;%%
}

\lref\GregoryTE{
  R.~Gregory, J.~A.~Harvey and G.~W.~Moore,
  ``Unwinding strings and T-duality of Kaluza-Klein and H-monopoles,''
  Adv.\ Theor.\ Math.\ Phys.\  {\bf 1}, 283 (1997)
  [arXiv:hep-th/9708086].
  %%CITATION = HEP-TH 9708086;%%
}

\lref\gssy{D.~Gaiotto, A.~Simons, A.~Strominger and X.~Yin,
``D0-branes in Black Hole Attractors,'' arXiv:hep-th/0412179.}

\lref\eemr{
  H.~Elvang, R.~Emparan, D.~Mateos and H.~S.~Reall,
  ``A supersymmetric black ring,''
  Phys.\ Rev.\ Lett.\  {\bf 93}, 211302 (2004)
  [arXiv:hep-th/0407065].
}

\lref\gv{R.~Gopakumar and C.~Vafa, ``M-theory and Topological
Strings - I,II", hep-th/9809187; hep-th/9812127.}

\lref\BenaTK{
  I.~Bena and P.~Kraus,
  ``Microscopic description of black rings in AdS/CFT,''
  JHEP {\bf 0412}, 070 (2004)
  [arXiv:hep-th/0408186].
  %%CITATION = HEP-TH 0408186;%%
}

\lref\jmjmas{\ J.~M.~Maldacena, J.~Michelson and A.~Strominger,
``Anti-de Sitter fragmentation,'' JHEP {\bf 9902}, 011 (1999)
[arXiv:hep-th/9812073]. }

\lref\juan{ J.~M.~Maldacena, `The large N limit of superconformal
field theories and supergravity,'' Adv.\ Theor.\ Math.\ Phys.\
{\bf 2}, 231 (1998) [Int.\ J.\ Theor.\ Phys.\  {\bf 38}, 1113
(1999)] [arXiv:hep-th/9711200]. }

\lref\GauntlettQY{
  J.~P.~Gauntlett and J.~B.~Gutowski,
  ``General concentric black rings,''
  Phys.\ Rev.\ D {\bf 71}, 045002 (2005)
  [arXiv:hep-th/0408122].
  %%CITATION = HEP-TH 0408122;%%
}

\lref\kall{R.~Kallosh, A.~Rajaraman, W.~K.~Wong, ``Supersymmetric
Rotating Black Holes and Attractors", Phys.\ Rev.\ D {\bf 55},
3246 (1997) [arXiv:hep-th/9611094].}

\lref\kkv{S.~Katz, A.~Klemm and C.~Vafa, ``M-Theory, Topological
Strings and Spinning Black Holes", Adv.\ Theor.\ Math.\ Phys.\
{\bf 3}, 1445 (1999) [arXiv:hep-th/9910181]. }

\lref\KrausGH{
  P.~Kraus and F.~Larsen,
  ``Attractors and black rings,''
  arXiv:hep-th/0503219.
  %%CITATION = HEP-TH 0503219;%%
}

\lref\MohauptMJ{ T.~Mohaupt, ``Black hole entropy, special
geometry and strings,'' Fortsch.\ Phys.\  {\bf 49}, 3 (2001)
[arXiv:hep-th/0007195]. }

\lref\fivedsugra{M.~Gunaydin, G.~Sierra and P.~K.~Townsend,
  ``Gauging The D = 5 Maxwell-Einstein Supergravity Theories: More On Jordan Algebras,''
  Nucl.\ Phys.\ B {\bf 253}, 573 (1985).}

\lref\msw{ J. Maldacena, A. Strominger and E. Witten, ``Black Hole
Entropy in M-Theory", hep-th/9711053. }

\lref\osv{ H.~Ooguri, A.~Strominger and C.~Vafa, ``Black hole
attractors and the topological string,'' arXiv:hep-th/0405146.
%%CITATION = HEP-TH 0405146;%%
}

\lref\rgcv{ R.~Gopakumar and C.~Vafa, ``M-theory and topological
strings. I,'' arXiv:hep-th/9809187. }

\lref\shmakova{ M.~Shmakova, `Calabi-Yau black holes,'' Phys.\
Rev.\ D {\bf 56}, 540 (1997) [arXiv:hep-th/9612076].
%%CITATION = HEP-TH 9612076;%%
}

\lref\SimonsNM{ A.~Simons, A.~Strominger, D.~M.~Thompson and
X.~Yin, ``Supersymmetric branes in AdS(2) x S**2 x CY(3),''
arXiv:hep-th/0406121.
%%CITATION = HEP-TH 0406121;%%
}

\lref\spinor{T. Mohaupt, ``Black Hole Entropy, Special Geometry
and Strings", hep-th/0007195.}

\lref\ssty{ A.~Simons, A.~Strominger, D.~M.~Thompson and X.~Yin,
``Supersymmetric branes in AdS(2) x S**2 x CY(3),''
arXiv:hep-th/0406121. }

\lref\suss{ B.~Freivogel, L.~Susskind and N.~Toumbas, ``A two
fluid description of the quantum Hall soliton,''
arXiv:hep-th/0108076.
%%CITATION = HEP-TH 0108076;%%
}

\lref\VafaGR{ C.~Vafa, ``Black holes and Calabi-Yau threefolds,''
Adv.\ Theor.\ Math.\ Phys.\  {\bf 2}, 207 (1998)
[arXiv:hep-th/9711067].
%%CITATION = HEP-TH 9711067;%%
}

\lref\VerCK{
  E.~Verlinde,
  ``Attractors and the holomorphic anomaly,''
  arXiv:hep-th/0412139.
}

\lref\GaiottoGF{
  D.~Gaiotto, A.~Strominger and X.~Yin,
  ``New Connections Between 4D and 5D Black Holes,''
  arXiv:hep-th/0503217.
}

\lref\CyrierHJ{
  M.~Cyrier, M.~Guica, D.~Mateos and A.~Strominger,
  ``Microscopic entropy of the black ring,''
  arXiv:hep-th/0411187.
}

 \Title{\vbox{\baselineskip12pt\hbox{hep-th/0504126} }}{{ 5D
Black Rings and 4D Black Holes}}

\centerline{Davide Gaiotto\footnote{*}{Permanent address:
Jefferson Physical Laboratory, Harvard University, Cambridge, MA,
USA.} ,~ Andrew Strominger* and Xi Yin* }
\smallskip\centerline{Center of Mathematical Sciences}
\centerline{ Zhejiang University, Hangzhou 310027 China}

\centerline{} \vskip.6in \centerline{\bf Abstract} { It has
recently been shown that the M theory lift of a IIA 4D BPS
Calabi-Yau black hole is a 5D BPS black hole spinning at the
center of a Taub-NUT-flux geometries, and  a certain linear
relation between 4D and 5D BPS partition functions was accordingly
proposed. In the present work we fortify and enrich this proposal
by showing that the M-theory lift of the general 4D multi-black
hole geometry are 5D black rings in a Taub-NUT-flux geometry.

 } \vskip.3in

\smallskip
\Date{}

\listtoc \writetoc

\newsec{Introduction}

In a recent paper \GaiottoGF\ it was shown that the M-theory lift
of a general 4D BPS Calabi-Yau black hole can be viewed as a 5D
BPS spinning black hole \bmpv\  sitting in the center of a
flux-Taub-NUT spacetime. This suggested a direct relation between
the microstates of 4D and 5D black holes, and (invoking \osv)
motivated a conjecture of the form
\eqn\guj{Z_{5D}=Z_{4D}=|Z_{top}|^2,} relating a 4D BPS partition
function, a 5D BPS partition function and the topological string.
See \GaiottoGF\ for the precise form of and arguments appearing in
\guj.

A recent surprise \eemr\ is that BPS objects in 5D include black
rings as well as black holes. The 5D partition function $Z_{5D}$
should include all BPS states, in particular black rings. A check
of our conjecture is then that $Z_{4D}$ must include a
contribution to match the 5D black rings in $Z_{5D}$.

Indeed we will show that there is just such a matching
contribution. In a series of beautiful papers
\refs{\denefb,\denefc,\BatesVX} Denef and collaborators have
explicitly constructed multi-center 4D BPS black hole solutions
which in general carry angular momenta. The black holes in these
solutions can have different sets of charges and they are bound to
one another in the sense that the black holes separations are
fixed in terms of their charges and the asymptotic values of the
moduli. In this paper we will construct exact
"flux-Taub-NUT-black-ring" solutions describing a black ring in
Taub-NUT with four-form flux turned on. We further show that these
solutions are precisely the lift to 5D of the Denef multi-center
solutions, and the 4D black hole separations become the radii of
the 5D rings. This result fortifies and enriches the 4D-5D
connection proposed in \GaiottoGF,  and will also clearly have
implications for our understanding of the topological
string.\foot{In particular we expect the anomalous background
dependence of the topological string to be equivalent to the
anomalous asymptotic moduli dependence of the black hole partition
function arising from split attractor flows as analyzed in
\refs{\MooreZU,\MoorePN,\MooreFG,\BatesVX,\denefb,\denefc}.}

This paper is organized as follows. Section 2 reviews the
multi-center black hole solutions. Section 3 constructs a general
solution of 5D supergravity describing black rings and black holes
in a multi-Taub-NUT-flux geometry.  In section 4 we show that
section 3 is the M-theory lift of section 2, and describe the
basic example of lifting the bound state of a D6-brane with a
D4-D2-D0 black hole to a 5D black ring.

\newsec{Multicenter BPS solutions in 4D}

In this section we review, and slightly reformulate, the general
multi-center BPS solution of 4D ${\cal N}=2$ supergravity\foot{The
solutions reviewed here solve the leading order equations,and do not
incorporate $R^2$ corrections.} found in \BatesVX. The solution is
characterized by electromagnetic charges and asymptotic moduli. It
may be expressed in terms of $2h_{11} + 2$ real harmonic functions
on $R^3$ \eqn\ab{\eqalign{ {H^\Lambda(\vec x) \choose H_\Lambda(\vec
x)} = {h^\Lambda \choose h_\Lambda} + \sum_s {p^\Lambda_i \choose
q_{\Lambda,i}}{1 \over |\vec x - \vec x_i|} } ,} where
$(p^\Lambda_i,q_{\Lambda i})$ is the electromagnetic charge located
at the spatial position $\vec x_i$, and $(h^\Lambda,h_\Lambda)$ are
constants which will shortly be related to the asymptotic moduli.
The projective scalar moduli $X^\Lambda$ as a function of spatial
positions are then given by \eqn\attr{\eqalign{ CX_{}^\Lambda (\vec
x)= H^\Lambda(\vec x) + {i\over \pi} \left.{{\partial
S_{bh}(p^\Lambda,q_{\Lambda})}\over{\partial
q_\Lambda}}\right|_{p^\Lambda=H^\Lambda(\vec
x),~q_{\Lambda}=H_\Lambda(\vec x)} }.}  The complex function $C$
depends on the choice of projective gauge (and may be set to one by
an appropriate choice). As a function of the moduli and prepotential
(or periods $F_\Lambda$) $S_{bh}$ here is given by
\eqn\rdf{S_{bh}(CX^\Lambda(\vec x))={\pi\over 2}{\rm Im}[CX^\Lambda
\bar C \bar F_\Lambda ].} In order to find $S_{bh}$ as a function of
charges, as needed in \attr,  one must solve the algebraic attractor
equations \refs{\fks,\aes}. This may or may not be analytically
possible, depending on the form of the prepotential and the charge
vector. Note that the $S_{bh}$ used here is a function of position
and is equal to the black hole entropy only at the horizon. The
constants $h$ encode the values of the moduli at spatial infinity,
i.e. $Re [CX^\Lambda(\infty)]=h^\Lambda,  Re
[CF^\Lambda(\infty)]=h_\Lambda$.

Given the moduli fields $X^\Lambda (\vec x)$ the four dimensional
metric is then simply \eqn\ac{\eqalign{ ds_4^2 = -{\pi\over S_{bh} }
\left(dt + \omega^{(4)}\right)^2 +{S_{bh}\over \pi} d\vec x^2 } ,}
where it is implicit that $S_{bh}=S_{bh}(X^\Lambda(\vec x))$ and
$\omega^{(4)}$ is the solution of \eqn\sar{d\omega^{(4)}= H_\Lambda
*^3 dH^\Lambda - H^\Lambda *^3 dH_\Lambda.} The gauge fields
strengths are\eqn\acc{dA^\Lambda = d \left[ S_{bh}^{-1}{\partial
S_{bh}\over{\partial q_\Lambda} }\left(dt +
\omega^{(4)}\right)\right]_{p^\Lambda=H^\Lambda,~q_{\Lambda}=H_\Lambda}
+ *^3 d  H^\Lambda .}

Finally the equilibrium positions $\vec x_j$ of the black hole
centers are determined by the asymptotic moduli and the charges
via the integrability condition following from \sar, which may be
written \eqn\ad{ \left.\left[p_j^\Lambda H_\Lambda(\vec x) -
q_{j\Lambda }H^\Lambda(\vec x) \right]\right|_{\vec x = \vec x_j}
= 0 .}

The basic example of this paper, which illustrates the connection
to the black ring, is a bound state of a single $D6$ brane at
$r=0$ and a $D4-D2-D0$ black hole with charges $(p^A,q_A,q_0)$ at
$r=L, ~\theta=0$. The harmonic functions are:

\eqn\ae{\eqalign{ &H^0 = {4\over R_{TN}^2} + {1\over r} \cr &H^A =
{p^A\over (r^2 + L^2 - 2 r L \cos \theta)^{1\over 2}} \cr &H_A = h_A
+ {q_A \over (r^2 + L^2 - 2 r L \cos \theta)^{1\over 2}} \cr & H_0 =
- {q_0\over L} + {q_0\over (r^2 + L^2 - 2 r L \cos \theta)^{1\over
2}} } } The integrability condition \ad\ is then \eqn\hds{{1\over L}
+ {4\over R_{TN}^2} = {h_A p^A\over q_0}.} As the parameter $R_{TN}$
goes to infinity the distance between the centers reaches a minimum
value, while for $R_{TN}$ small enough the distance between the
centers will diverge, and the bound state disappears.

\newsec{5D solutions}

This section will describe some new supersymmetric 5D black
ring-Taub-NUT-flux  solutions which generalize previous solutions
of minimal supergravity \refs{\eemr,\KrausGH}. In the next section
we will see they are simply the lift to 5D of the 4D multicenter
solutions reviewed in the previous section.

${\cal N}=2$ supergravity fields in 5D are organized by the
so-called very special geometry \fivedsugra, parameterized by
$h_{11}$ real scalar fields $Y^A$, subject to the constraint
\eqn\cdt{D_{ABC} Y^A Y^B Y^C = 1,} for constant couplings $D_{ABC}$.
It is useful to further define \eqn\wsz{Y_A \equiv 3D_{ABC} Y^B Y^C
.} BPS solutions in 5D ${\cal N} = 2$ supergravity may be written
following \refs{\allfive, \GutowskiYV, \GauntlettQY, \KrausGH}
\eqn\ak{\eqalign{ & ds_5^2 = -2^{-4/3}f^{-2} \left( dt + \omega
\right)^2 + 2^{2/3}f ds_{X}^2 \cr &F^A = d\left[f^{-1}Y^A(dt +
\omega)\right] + \Theta^A } } where $ds_{X}^2$ is hyperk\"{a}ler
metric  on a 4D hyperk\"{a}ler space $X$,  $\Theta^A$ are closed
self-dual 2-forms on $X$, the self-dual part of $d\omega$ is $- f
Y_A \Theta^A$  and $f$ is a function on $X$ obeying \eqn\al{
\nabla^2 (f Y_A) = {3} D_{ABC} \Theta^B \cdot \Theta^C .} When the
space $X$ is Taub-NUT,\foot{The more general solution with $X$ being
a Gibbons-Hawking space was presented in \GauntlettQY.} one has
\eqn\am{ ds_X^2 = H^0(\vec x) d\vec x^2 + H^0(\vec x)^{-1} (dx^5 +
\omega^0)^2,~~~~ d\omega^0 = *^3 dH^0 ,} with $H^0$ a harmonic
function as in \ae, and the coordinate $x^5$ has periodicity $4\pi$.
Closed self-dual 2-forms are then given by \eqn\wse{\Theta^A =
d\left[{H^A\over H^0} (dx^5 + \omega^0)\right] + *^3 dH^A ,} with
harmonic $H^A$ as in \ae. Inserting \wse\ the equation \al\ for $f$
becomes \eqn\hgf{ \nabla^2 (f Y_A) = 6D_{ABC}\nabla \left({H^B\over
H^0}\right)\cdot\nabla \left({H^C\over H^0}\right).} This is
magically solved by \eqn\mkj{f Y_A = H_A + {3D_{ABC}H^B H^C\over
H^0} .} $f$ and $Y_A$ are then determined by \cdt\ and \mkj. This
immediately gives the self-dual part of $d\omega$. It is
straightforward to show that $\omega$ can be written
\eqn\fgm{\omega=-\left(H_0 + 2 {D_{ABC} H^A H^B H^C \over (H^0)^2} +
{H_A H^A \over H^0}\right)(dx^5+\omega^0)+\omega^{(4)},} where
$\omega^{(4)}$ satisfies the integrability condition \ad.

As we will see more clearly in the next section, the solution \ak\
together with \wse,\mkj,\fgm\ describe black rings in the Taub-NUT
space.

\newsec{4D $\to$ 5D lift}
${\cal N}=2$ 5D supergravity may be viewed as the circle
decompactification of ${\cal N}=2$ 4D supergravity. In this
section we see that the 5D black ring solution of the previous
section is the lift from 4D of the general multi-center black hole
solutions. A discussion of the relevant geometry is in \GregoryTE.

Given a solution of $ds_{4D}^2,A_{4D}^A, A_{4D}^0, z^A={X^A \over
X^0}$ of 4D supergravity, a solution of 5D supergravity is quite
generally given by \eqn\aa{ \eqalign{ ds_{5D}^2 &= 2^{2/3}{\cal
V}^2(dx^5 + A_{4D}^{0})^2+ 2^{-1/3}{\cal V}^{-1} ds_{4D}^2, \cr
A_{5D}^{A} &= A_{4D}^{A} + {\rm Re} z^A (dx^5 + A_{4D}^{0}), \cr
Y^A&={\cal V}^{-1}{\rm Im} z^A ,~~~~ ~~~~~~~{\cal V}\equiv
\left(D_{ABC}{\rm Im} z^A{\rm Im} z^B{\rm Im} z^C\right)^{1\over
3}.} } Inserting the multi-center 4D solution of section 2 then
gives the general black ring solutions of section 3.

To be more explicit, let us consider the 4D prepotential \eqn\af{
F(X) = {D_{ABC}X^A X^B X^C \over X^0} .} The expression for the
entropy as a function of the charges is known, although complicated
\shmakova. It is \eqn\ag{\eqalign{ &S(p,q) = 2\pi\sqrt{Q^3 p^0 - J^2
(p^0)^2},\cr &Q^{3\over 2} = D_{ABC} y^A y^B y^C , \cr &3 D_{ABC}y^A
y^B  = q_C + {3D_{ABC} p^A p^B \over p^0}, \cr &J = {q_0\over 2} +
{D_{ABC} p^A p^B p^C \over (p^0)^2} + {p^A q_A\over 2p^0}. } }
Correspondingly there will be certain functions $Q(\vec x)$ and
$J(\vec x)$ built out of the harmonic functions $(H^\Lambda(\vec
x),H_\Lambda (\vec x))$. The volume of the Calabi-Yau at $\vec x$ is
${\cal V}(\vec x)^3$, with \eqn\volcy{ {\cal V}(\vec x) =
{S(H^\Lambda, H_\Lambda)\over 2\pi H^0 Q} } With a bit algebra, \aa\
then yields for the metric \eqn\ah{\eqalign{ & ds_{5D}^2 =
-2^{-4/3}Q(\vec x)^{-2} \left( dt + \omega^{(4)} -2 J(\vec x)(dx^5 +
\omega^0) \right)^2 +2^{2/3}Q(\vec x) ds_{TN}^2 \cr & ds_{TN}^2 =
H^0(\vec x) d\vec x^2 + H^0(\vec x)^{-1}(dx^5 + \omega^0)^2 } }
where \eqn\dfe{d\omega^0=*^3dH^0} and $\omega^{(4)}$ is given by
\sar. This is precisely the solution we found in section 3. The 4D
multi-centered black hole solution in general has several horizons,
which lifts to horizons in 5D. These horizons are circle bundles
over $S^2$. Depending on whether each center has nonzero magnetic KK
charge, the 5D horizons will be either quotients of $S^3$,
corresponding to a 5D spinning black hole, or $S^2\times S^1$,
corresponding to a black ring.

Now let us further specialize to the bound state of a single
D6-brane with a D4-D2-D0 black hole with charges $(p^A,q_A, q_0)$.
The relevant harmonic functions are given in \ae. In the limit
$R_{TN}\to\infty$, $H^0 \to {1 \over r}$, \ah\ becomes precisely the
black ring solution in flat 5D spacetime \eemr. The radius of the
ring is $R_{ring}=L$. It is constructed from wrapped M5 branes with
charges $p^A$, carries M2 charges $\tilde q_A=q_A+3D_{ABC}p^Bp^C$
and $SU(2)_L$ spin $J_L=q_0/2$.

Note that the entropy of the two-centered black hole comes from only
the D4-D2-D0 system. It is amusing to verify directly that the tree
level entropy of the D4-D2-D0 system of charge $(p^A, q_A, q_0)$
indeed agrees with that of the black ring \BenaTK,\CyrierHJ\ with
M5-M2 charge $(p^A, \tilde q_A)$ and angular momentum $J_L=q_0/2$.

More generally, when there are mulitple 4D black holes carrying D6
charge, the background geometry of the 5D lift will be a resolved
multi-Taub-Nut-flux geometry.  The black holes of the 4D solution
that carry $D6$ charges will lift to 5D spinning black holes at
the fixed points of the $U(1)_L$ isometry of  the multi-Taub-NUT
background. Those that do not will lift to 5D black rings tracing
orbits of the isometry.

\bigskip

\centerline{\bf Acknowledgements} We would like to thank M. Guica
for useful conversations. This work was supported in part by DOE
grant DE-FG02-91ER40654.

\listrefs

\end